\documentclass[twoside]{article}
\usepackage{multicol,amsmath}  

\evensidemargin=\oddsidemargin \oddsidemargin=0.5cm
\font\twbf=cmbx10 at 11pt   
\font\bsl=cmbxsl10 at 11pt 
\font\bfit=cmbxti10   

\catcode`\@=11
\newcounter{sectc}\newcounter{subsectc}\newcounter{subsubsectc}
\newcommand{\sect}[1] {\vspace{0.3cm}\addtocounter{sectc}{1} 
\setcounter{subsectc}{0}\setcounter{subsubsectc}{0}\noindent 
      {\twbf\thesectc\ \  #1}\par\vspace{0.2cm}}
\newcommand{\subsect}[1] {\vspace{0.3cm}\addtocounter{subsectc}{1} 
          \setcounter{subsubsectc}{0}\noindent 
          {\bf\thesectc.\thesubsectc\ \ \bfit  #1}\par\vspace{0.15cm}}

\newcommand{\nonumsect}[1] {\vspace{0.3cm}\noindent{\twbf #1}
          \par\vspace{0.2cm}}
\newcounter{appendixc}
\newcounter{subappendixc}[appendixc]
\newcounter{subsubappendixc}[subappendixc]

\renewcommand{\appendix}[1] {\vspace{0.3cm}
        \refstepcounter{appendixc}
        \setcounter{figure}{0}
        \setcounter{table}{0}
        \setcounter{equation}{0}
        \renewcommand{\thefigure}{\Alph{appendixc}.\arabic{figure}}
        \renewcommand{\thetable}{\Alph{appendixc}.\arabic{table}}
        \renewcommand{\theappendixc}{\Alph{appendixc}}
        \renewcommand{\theequation}{\Alph{appendixc}.\arabic{equation}}
        \noindent{\bf Appendix \theappendixc\ \ #1}\par\vspace{0.2cm}}

\if@twoside \def\ps@headings{\let\@mkboth\markboth
\def\@oddfoot{}\def\@evenfoot{}\def\@evenhead{\rm\footnotesize \thepage\hfil 
\footnotesize
\leftmark}\def\@oddhead{\hbox{}\footnotesize \rightmark \hfil
\rm\thepage}\def\sectionmark##1{\markboth {\uppercase{\ifnum \c@secnumdepth
>\z@
 \thesection\hskip 1em\relax \fi ##1}}{}}\def\subsectionmark##1{\markright
{\ifnum \c@secnumdepth >\@ne
 \thesubsection\hskip 1em\relax \fi ##1}}}
\else \def\ps@headings{\let\@mkboth\markboth
\def\@oddfoot{}\def\@evenfoot{}\def\@oddhead{\hbox {}\footnotesize \rightmark 
\hfil\rm\footnotesize\thepage}\def\sectionmark##1{\markright 
{\uppercase{\ifnum \c@secnumdepth
>\z@
 \thesection\hskip 1em\relax \fi ##1}}}}
\fi
\def\ps@myheadings{\let\@mkboth\@gobbletwo
\def\@oddhead{\hbox{}\footnotesize\rightmark \hfill\rm\footnotesize\thepage}
\def\@oddfoot{}
\def\@evenhead{\rm\footnotesize\thepage 
 \hfill \footnotesize\leftmark\hbox{}}\def\@evenfoot{}
\def\sectionmark##1{}\def\subsectionmark##1{}}
\def\d{\hspace*{.8\p@}\mathstrut\text{d}\hspace*{.6\p@}} 
\def\e{\mathstrut\,\text{e\/}\hspace*{.6\p@}} 
\def\i{\hspace*{.6\p@}\mathstrut\text{i}\hspace*{.6\p@}} 
\catcode`\@=13
\def\Td#1#2#3#4#5{{\thispagestyle{empty}
\protect\headheight0pt\protect\headsep0pt\protect\vspace*{-2.2cm}
{\flushleft\parbox{181mm}{\footnotesize Commun.\ Theor.\ Phys.~(Beijing, 
China) {\bf #1}{~(#2)~}{pp~#3}\\[-0.7mm]
\copyright\hspace*{3.5pt} International Academic Publishers\hfill Vol.~{#4},
No.~{#5}}\\[-1.4mm]
\begin{table}[h]\hfill\null\hfill\hrule\vskip.4mm\hrule\end{table} }}}
\def\no#1{\rlap{\protect\rule[-0.25 true cm]{\textwidth}{0.03 true cm}}%
No.~{#1}\hfill}
\def\vo#1{\hfill{\ignorespaces Vol.~{#1}%
\llap{\protect\rule[-0.25 true cm]{\textwidth}{0.03 true cm}}}}

\def\rd{\protect\footnotesize}
\def\pacs#1{\protect\vglue6pt%
\noindent\begin{minipage}{170mm}{\bf PACS numbers: }\rm #1\end{minipage}}
\def\key#1{\leftskip=0.5cm\vglue0pt\noindent\hspace*{-1pt}\protect
\begin{minipage}{170mm}\begin{minipage}[t]{21mm}{\bf Key words:}\end{minipage}
\hfill\begin{minipage}[t]{148mm}\rm\baselineskip=12pt #1\end{minipage}
\end{minipage}\par\leftskip=0cm}
\def\title#1{\begin{flushleft} 
\large\bf\protect\baselineskip=17pt #1\end{flushleft}}
\def\author#1{\leftskip=0.5cm\noindent\begin{minipage}{170mm}
\normalsize\hspace*{-4.5pt}#1\end{minipage}\par\vglue4pt} 
\def\address#1#2{\leftskip=0.5cm\noindent\begin{minipage}{170mm}\parindent=-4.5pt   
${}^{#1}$\protect\small\baselineskip=10pt #2\end{minipage}\par\vglue2pt} 
 
\def\date#1{\leftskip=0.5cm\vglue4pt\noindent\begin{minipage}{170mm}
\normalsize\hspace*{-4.5pt}#1\end{minipage}\par\vglue4pt} 
\def\abstract#1#2{\leftskip=0.5cm\vglue4pt\noindent\hspace*{-4.5pt}
\begin{minipage}{170mm}\small\sl\baselineskip=11pt{\bsl Abstract\ } #1\par
\pacs{\normalsize#2}\end{minipage}\par\vglue2pt\leftskip=0cm} 
    
\def\ruleup{\vspace*{-0.5cm}\noindent\rule{9.05cm}{0.6pt}\rule{0.6pt}{0.4cm}}
\def\ruledown{\hfill\noindent{\lower.38cm\hbox{\rule{0.6pt}{0.4cm}}\rule{9.05cm}
              {0.6pt}}\vspace*{-0.6cm}}
\def\refrule{\vglue5pt\centerline{\hbox to 10cm{\hrulefill}}\vglue10pt}
\def\wen#1{$^{[#1]}$}
\def\noa#1{\noalign{\vskip#1pt}}
\def\alw{\allowdisplaybreaks}
\def\ssc#1{\scriptscriptstyle{#1}}

\begin{document}

\hoffset=-1.4cm
\voffset=0.4cm

\abovedisplayskip=6pt plus 1pt minus 1pt
\belowdisplayskip=6pt plus 1pt minus 1pt
\parskip=0.1pt plus.3pt minus0.1pt

\setcounter{footnote}{0}
\setcounter{page}{703} 
\Td{35}{2001}{703--710}{35}{6, June 15, 2001} 

\def\om{\omega} \def\we{\wedge} \def\va{\varepsilon} \def\omb{\bar{\omega}}
\def\la{\lambda} \def\vv{\f{V}{\la^d}} \def\ep{\epsilon}
\def\Si{\Sigma} \def\si{\sigma} \def\t{T_+} \def\bi{\bibitem}
\def\c{\cite} \def\D{\Delta} \def\a{\alpha} \def\b{\beta} \def\pa{\partial}

\title{On Symplectic and Multisymplectic Structures and Their Discrete
Versions in Lagrangian Formalism}
\author{GUO Han-Ying, LI Yu-Qi and WU Ke}
\address{}{Institute of Theoretical Physics, Academia Sinica,
P.O.\ Box 2735, Beijing 100080, China}
\date{(Received April 2, 2001)}
\abstract{We introduce the Euler--Lagrange cohomology to study the 
symplectic and multisymplectic structures and their 
preserving properties in finite and infinite dimensional Lagrangian 
systems  respectively.  We also explore their certain difference discrete 
counterparts in the relevant regularly discretized 
finite and  infinite dimensional Lagrangian systems by
means of the difference discrete variational principle with the difference
being regarded as an entire geometric object and the noncommutative 
differential calculus  on regular lattice. In order to show that in 
all these cases the symplectic and multisymplectic
preserving properties do not necessarily depend on the relevant 
Euler--Lagrange 
equations, the Euler--Lagrange cohomological concepts and content in the 
configuration  space are employed.}{11.10.Ef, 02.60.Lj}
\key{Euler--Lagrange cohomology, difference discrete variational
principle, symplectic structure}


\vspace*{0.4cm}

\baselineskip=13pt plus.4pt minus.2pt

\begin{multicols}{2}
\sect{Introduction}

It is well known that the symplectic and multisymplectic structures
play  crucially important roles in the symplectic and multisymplectic
algorithms for the finite dimensional Hamiltonian systems\wen{1,2}
and infinite-dimensional Hamiltonian systems respectively. 
These algorithms are  very powerful  and successful  in  numerical   
calculations of  the relevant  systems in comparison with  other 
various non-symplectic computational  schemes since
the  symplectic and multisymplectic schemes   preserve the symplectic 
structure and multisymplectic structure of the systems in certain sense. 
 
In this paper, in a simple and direct manner, we present the symplectic 
structure in the Lagrangian mechanism and the multisymplectic 
structure in the Lagrangian field theory and their preserving properties 
as well as the difference-type discrete versions of these issues. We employ
the ordinary exterior differential calculus in the configuration space 
and introduce what is named the Euler--Lagrange (EL) cohomology to show 
that the symplectic and multisymplectic  structures are preserved without 
necessarily  making use of the EL equations in general. And the EL equations 
are derived from the variational principle of the relevant action functionals.
Therefore, it is important to emphasize that these structure-preserving 
properties are established in the function space with the EL cohomology on the 
configuration space in general  rather than in the solution space of the 
EL equations only. 

One of the key points different from the other approaches in our approach 
  is the EL cohomology we will introduced. Some EL cohomological 
concepts and content such as the EL one-forms, the coboundary EL one-forms
 and the EL conditions, i.e., the EL one-forms are closed,  in each case will
play  important roles. In each case, the null EL one-form gives rise to the 
EL equations. Although the null EL one-form is a special case of the 
coboundary EL one-forms which are cohomologically trivial and automatically 
satisfy the (closed) EL condition,   they are {\bfit not} the 
same in principle. As a matter of fact, the EL cohomology is nontrivial 
in general. This point plays a crucial role in our approach. 

In the  course of numerical calculation, the ``time''  $t \in R$ is always 
discretized, say, with equal spacing $h=\Delta t$ and the space coordinates 
are also discretized in various cases, especially, for the Lagrangian
field theory. In addition to these computational
approach, there  also exist various discrete physical systems with 
discrete or difference discrete Lagrangian functions.
It is well known that the differences of functions 
do not obey the Leibniz law. In order to explore  the discrete 
symplectic and multisymplectic  structures in these difference discrete 
systems and their preserving properties in certain difference discrete 
versions,  some noncommutative differential calculus (NCDC) should be 
employed\wen{5-7} even for the well-established symplectic algorithms. 
This is the second key point of this paper. 

Another key point of this paper different from others is that the difference 
discrete variational  principle (DDVP) will be employed in this paper. 
In view of NCDC, a forward or backward difference as
 the forward or backward discrete derivative should
be regarded as an entire geometric object respectively. In the DDVP with 
forward (or backward) difference, we  prefer to adopt this point of view. 
We also show that DDVP leads to the correct results in the sense that 
the results not only correspond to  the correct
ones in the continuous limit but also are the wanted discrete version 
in various well-known cases.

The plan of this paper is as follows. We first briefly rederive some 
well-known contents on symplectic and multisymplectic structures and 
their preserving properties in the
Lagrangian formalism for the finite  and infinite dimensional 
 systems respectively in Sec.~2. The important issues of this section is 
to introduce the  EL cohomology, including some cohomological 
concepts and content such as the EL one-forms, the coboundary EL one-forms
and the EL conditions and to show that it is nontrivial in each case. 
In order to explain those symplectic and multisymplectic geometry and 
relevant preserving properties in relevant systems the EL cohomology 
plays a very important role. In Sec.~3, we first 
explain the DDVP in our approach to give rise to the discrete 
Euler--Lagrange (DEL) equations. Then we study the difference discrete 
versions of  the cohomological concepts and content as well as the symplectic 
and multisymplectic  structures in Lagrangian formalism given in 
Sec.~2. We  present some remarks in Sec.~4. Finally, in the Appendix, 
some relevant NCDC on regular lattice with equal spacing on each direction is
given.

\sect{The Symplectic and Multisymplectic Struc-\newline\phantom{2\ \
}tures in Lagrangian Mechanism and Field\newline\phantom{2\ \ }Theory}
\vspace*{-0.15cm}

In this section, we  recall some well-known contents on
symplectic and multisymplectic structures and their preserving in the
Lagrangian formalism for the finite  and infinite dimensional 
 systems respectively. The important point is to introduce the EL 
cohomological concepts and content related to
the EL equations  and explain their important roles in the symplectic
and multisymplectic geometry in these systems respectively.

\subsect{The Symplectic Structure in Lagrangian\newline\phantom{2.1\
\ }Mechanism} 

We begin with the Lagrangian mechanism.  Let time $t\in R^1$ be the base 
manifold, $M$ the configuration space  on $t$ with coordinates $q^i(t)$, 
$i=1, \ldots, n$, $TM$ the tangent bundle of $M$ with coordinates 
$q^i, \dot q^j$,   $F(TM)$ the function space on $TM$. 

The Lagrangian of the systems is denoted as $L(q^i, {\dot q^j})$, where 
$\dot q^j$ is the derivative of $q^j$ with respect to $t$. The variational 
principle gives rise to the well-known EL equations  
\begin{align}
\frac{\partial L}{\partial {q^i}}-\frac{\d}{\d t} 
\frac {\partial L} {\partial {\dot q^i}}=0\,.
\end{align}
Let us  take the exterior derivative d of the Lagrangian function, 
we get
$$
\d L=\Bigl\{\frac{\partial L}{\partial {q^i}}
-\frac{\d}{\d t}\frac{\partial L} {\partial {\dot q^i}}\Bigr\}
\d q^i +\frac{\d}{\d t}\Bigl\{\frac{\partial L}{\partial{\dot q^i}} 
\d q^i\Bigr\}\,. 
$$
Defining the EL one-forms on $T^*M$, 
\begin{align}
E(q^i, \dot q^i):\;=\Bigl\{\frac{\partial L}{\partial {q^i}}
-\frac{\d}{\d t}\frac{\partial L}{\partial{\dot q^i}}\Bigr\}\d q^i\,,
\end{align}
we have
\begin{align} 
\d L(q^i, \dot q^i)=E(q^i, \dot q^i)+\frac{\d}{\d t} \theta\, ,
\end{align}
where  $\theta$ is the canonical one-form
\begin{align}
\theta =\frac{\partial L}{\partial{\dot q^i} }\d q^i\,.
\end{align}
It is easy to see from the definition (2) and Eq.~(3)
 that first the null EL one-form is corresponding to the EL 
equation, secondly the null EL one-form is a special case of the coboundary EL
 one-forms,
\begin{align}
 E(q^i, \dot q^j)=\d\alpha (q^i, \dot q^j)\,,
\end{align}
where $\alpha (q^i, \dot q^j)$ is an arbitrary smooth function of
$ (q^i, \dot q^j)$, and thirdly the EL one-forms are not coboundary in general 
since $\theta$ is not a coboundary so that the EL cohomology is not trivial.

Making use of nilpotency of d on $T^*M$, $\d^2L(q, {\dot q})=0$,
it is easy to show that if and only if the EL one-form is closed with respect 
to d, which may be named the EL condition, i.e.,
\begin{align}
\d E(q^i, \dot q^j)=0\,,
\end{align}
the symplectic  conservation law with respect to $t$ follows
\begin{align}
\frac{\d}{\d t} \omega = 0\,,
\end{align}
where the symplectic structure $\om$ is given by
\begin{align}
\omega = \d\theta =\frac{\partial^2 L}{\partial{\dot q^i}{\partial q^j}} 
\d q^i \wedge \d q^j
+\frac {\partial^2 L}{\partial{\dot q^i}{\partial {\dot q^j}}} 
\d q^i \wedge \d{\dot q}^j\,.
\end{align}

It is important to note that although the null EL one-form 
and the coboundary EL 
one-forms satisfy the EL condition, it does not mean that the closed EL 
one-forms can always be exact. Namely, as  mentioned above, the EL 
cohomology is not trivial so that   the EL condition is {\bfit  not} 
cohomologically equivalent to the null EL one-form  or the 
coboundary EL one-forms
in general. This  also means that $q^i(t)$, $i=1, \ldots, n$ in the EL 
condition are  {\bfit not} in the solution space of the EL equations only. 
In fact,  they are still in the function space in general. Therefore, 
the symplectic two-form $\omega$  is conserved not 
only in the solution space of the equations but also  in the 
function space in general with respect to the duration of $t$ if and 
only if the EL condition is satisfied.

In order to transfer to the Hamiltonian formalism, we introduce conjugate 
momentum
\begin{align}
p_j=\frac{\partial L} {\partial \dot q^j}\,,
\end{align}
and take a Legendre transformation to get the Hamiltonian function
\begin{align}
H(q^i,p_j)=p_k \dot q^k -L(q^i, \dot q^j)\,.
\end{align}
Then the EL equations become the canonical equations as follows:
\begin{align}
\dot q^i=\frac {\partial H} {\partial {p_i}},\quad\quad
\dot p^j=-\frac {\partial H} {\partial { q^j}}\,.
\end{align}
It is clear that a pair of the EL one-forms should be introduced now 
\begin{align}
&E_1(q^i, p_j)=\Bigl(\dot q^j - \frac{\partial H}{\partial p_j}\Bigr)\d p_j\,,
\nonumber\\\noa2
& E_2(q^i, p_j)=\Bigl(\dot p_j +\frac{\partial H}{\partial q^j}\Bigr)\d q^j\,.
\end{align}
In terms of $z^T={(q^i, \ldots, q^n, p_1, \ldots, p_n)}$, the canonical 
equations and the EL one-form become
\begin{align}
&\dot z=J^{-1}\nabla_z H\,, 
\\\noa2
&E(z)=\d z^T(Jz-\nabla_z H)\,.
\end{align}
Now it is straightforward to show that the symplectic structure preserving law
\begin{align}
\frac{\d}{\d t}\omega=0,\qquad \omega=\d z^T \wedge J\d z
\end{align}
holds {\it if and only if} the (closed) EL condition is satisfied
\begin{align}
\d E(z)=0\,.
\end{align}

\subsect{The Multisymplectic Structure in
Lagrangian\newline\phantom{2.2\ \ }Field Theory} 

We now consider the multisymplectic structure in Lagrangian field theory
for the scalar fields. Let $X$ be an $n$-dimensional base manifold with 
coordinates $x^{\mu}$, $\mu=1, \ldots, n$, $M$ the configuration space
 on $X$ with scalar field variables $u^i(x)$, $i=1, \ldots, s$,
$TM$ the tangent bundle of $M$ with coordinates $u^i, u_{\mu}^j$, $u_{\mu}^j
={\partial u^i}/{\partial x^{\mu}}$,
  $F(TM)$ the function space on $TM$.

The Lagrangian of the systems under consideration is $L(u^i, u_{\mu}^j)$
with the well-known EL equations  from the variational principle,
\begin{align}
\frac {\partial L}{\partial {u^i}}-
\frac {\partial} {\partial x^{\mu}} 
\frac {\partial L} {\partial {u_{\mu}^i}}=0\,.
\end{align}

Let us introduce the EL one-form in the function space $F(TM)$,
\begin{align}
E(u^i, u_{\mu}^j):\;=\Bigl\{\frac {\partial L}{\partial {u^i}}-
\frac{\partial}{\partial x^{\mu}}\frac {\partial L}{\partial{u_{\mu}^i}}
\Bigr\}\d u^i\,.
\end{align}
It is easy to see that the null EL one-form is corresponding to the EL 
equations and it is a special case for coboundary EL one-forms
\begin{align}
E(u^i, u_{\mu}^j)=\d\alpha (u^i, u_{\mu}^j)\,,
\end{align}
where $\alpha (u^i, u_{\mu}^j)$  is an arbitrary smooth function of 
$(u^i, u_{\mu}^j)$. Although they are cohomologically trivial but it is 
already seen that in the EL one-forms, $(u^i, u_{\mu}^j)$ are {\bfit not} in 
the solution space of the EL equations only rather they are in the 
function space in general. 

We now take the exterior derivative d of the Lagrangian. In terms of the 
EL one-form we get 
\begin{align}
\d L(u^i, u_{\mu}^j)=E(u^i, u_{\mu}^j)+\partial_{\mu}\theta^{\mu}\,,
\end{align}
where $\theta^{\mu}$ are the canonical one-forms
\begin{align}
\theta ^{\mu}= \frac {\partial L} {\partial  u_{\mu}^i } \d u^i\, .
\end{align}
From Eq.~(20), it is easy to see that the EL one-forms are not
coboundary in general since the canonical one-forms $\theta^{\mu}$ are not 
coboundary so that the EL cohomology is not trivial in general.

By virtue of the nilpotency of d, $\d^2L(u^i, u_{\mu}^j)=0$,
it is easy to prove that if and only if  the EL condition
is satisfied, i.e., the EL one-form is closed 
\begin{align}
\d E(u^i, u_{\mu}^j)=0\,,
\end{align}
the multisymplectic structure preserving (MSSP) property, i.e., the 
multisymplectic conservation or divergence free law follows
\begin{align}
\sum _{\mu=1}^n \frac{\partial}{\partial x^{\mu}}  \omega^{\mu} = 0\,,
\end{align}
where the symplectic structures $\om^{\mu}$ are given by
\begin{align}
\omega^{\mu} =\d\theta^{\mu} =
\frac{\partial^2 L}{\partial {u_{\mu}^i}{\partial u^j}} 
\d u^i \wedge \d u^j
+\frac {\partial^2 L}{\partial{ u_{\mu}^i}{\partial u_{\mu}^j}} 
\d u^i \wedge \d u_{\mu}^j\,.
\end{align}

Similar to the finite dimensional case, it is also important to note again 
that although the null EL one-form, the coboundary EL one-forms satisfy 
the EL condition, it does not mean that the closed EL one-forms can always be
exact as  pointed out above. In addition,  $u^i(x)$s in the EL condition 
are  {\bfit not} in the solution space of the EL equations only in general. 
Therefore, the MSSP law holds, i.e., the multisymplectic two-forms 
$\omega^{\mu}$  are conserved, not only in the solution space of the equations
but also in the function space with the closed EL condition in general.

\sect{The Discrete Symplectic and Multisymplec-\newline\phantom{3\ \ }tic 
Structures in Lagrangian Formalism}
\vspace*{-0.15cm}

Now we consider certain difference discrete versions of the symplectic 
and multisymplectic structures and their preserving properties 
in the Lagrangian mechanism and field theory studied in the last section.

\subsect{The Discrete Symplectic Structure 
in Discrete\newline\phantom{3.1\ \ }Lagrangian Mechanism}

Let us first  consider the  symplectic structure and its preserving 
in the Lagrangian mechanism in case that ``time'' $t$ is discretized while
the configuration space at each moment of $t$ is still continuous.

Let us assume, without loss any generality, that in the  course of numerical 
calculation,  the ``time''  $t \in R$ is discretized with equal spacing 
$h=\Delta t $,
\begin{align}
t\in R  \rightarrow t\in  {\it T_D}=\{ (t_k , t_{k+1}=t_k+h,  \quad k \in
Z)\}\,.
\end{align}
At the moment  $t_k$, the configuration space is $M_k^n \in M_{T_D}^{n}=
\{\cdots M_1^n \times \cdots \times M_k^n \cdots \}$, its 
coordinates are denoted by $q^i_{(k)}$.
 
The difference discrete Lagrangian can be written by 
\begin{align}
L_{D (k)}=L_D(q^i_{(k)}, q^i_{t (k)})\,,
\end{align}
where ${q_t^j}_{(k)}$ is (forward) difference of ${q^j}_{(k)}$ at 
$t_k$ defined by 
\begin{align}
\D_t q^j_{(k)}:\;=\frac {\partial } {\partial t} q^j_{(k)}=q^j_{t(k)}
=\frac 1 h \{{q^j}_{(k+1)}-{q^j}_{(k)}\}\,.
\end{align}
It is the (discrete) derivative on $T({\it T_D})$ in the sense of NCDC
of a regular lattice $L^1$ with equal spacing\wen{5} and the same
notation for it as in the continuous case has been employed. It should be
noted that in what follows the difference is always viewed as an entire 
geometric object and its dual $\d t$ is the base of $T^*({\it T_D})$. 

We now consider the DDVP of the action functional
$$
\displaylines{\hspace*{10mm}
S_D=\sum_{k \in Z}L_D(q^i_{(k)}, q^j_{t (k)})\,,
\hfill\cr\noa2
\hspace*{10mm}
\delta S_D=\sum_{k \in Z}\Bigl\{\frac{\partial L_{D(k)}}{\partial q^i_{(k)}} 
\delta q^i_{(k)}+\frac {\partial L_{D(k)}}{\partial q^i_{t(k)}} \delta 
q^i_{t(k)}\Bigr\}\,.
\hfill{(28)}\cr}
$$
By means of the modified Leibniz law with respect to $\Delta_t$\wen{5,6}
(see the Appendix), we have
\begin{align*}
&\Delta_t\Bigl(\frac {\partial L_{D(k-1)}} {\partial q^i_{t(k-1)}}\delta
q^i_{(k)}\Bigr)
\\\noa2
&\qquad=\frac{\partial L_{D(k)}}{\partial q^i_{(k)}}\delta q^i_{t(k)}
+\Delta_t\Bigl(\frac{\partial L_{D(k-1)}}{\partial q^i_{t(k-1)}}\Bigr)
\delta q^i_{(k)}\,.
\end{align*}
Therefore,
\begin{align*}
&\delta S_{D}=\sum_{k \in Z}\Bigl\{\frac {\partial L_{D(k)}} {\partial 
q^i_{(k)}} -\Delta_t\Bigl(\frac{\partial L_{D(k-1)}}{\partial 
q^i_{t(k-1)}}\Bigr)\Bigr\}^i_{(k)}
\\\noa2
&\phantom{\D S_{D}=}+\sum_{k \in Z}\Delta_t
\Bigl(\frac{\partial L_{D(k-1)}}{\partial 
q^i_{t(k-1)}}\delta q^i_{(k)}\Bigr)\,.
\end{align*}
Using the properties 
\setcounter{equation}{28}
\begin{align}
\sum_{k \in Z}\Delta_t f(t_k)=f(t_k)\Bigm|_{k=-\infty}^{k=+\infty}\,,
\end{align}
and assuming $\delta q^i_k$ satisfy 
$$
 \delta q^i(\pm \infty)=0\,,
$$
it follows the DEL equations
\begin{align}
\frac {\partial L_{D(k)}} {\partial q^i_{(k)}} 
-\Delta_t\Bigl(\frac{\partial L_{D(k-1)}}{\partial q^i_{t(k-1)}}
\Bigr)=0\,.
\end{align}

It is easy to see that for the case with difference discrete Lagrangian
\begin{align}
L_D^{(k)}({q^i}^{(k)}, \D_t {q^j}^{(k)})
=\frac 1 2 (\D_t {q^i}^{(k)})^2-V({q^i}^{(k)})\,,
\end{align}
the DDVP gives the DEL equations
$$
\Delta_t(\D_t {q^i}^{(k-1)})-V'({q^i}^{(k)})=0\,,
$$
i.e.,
\begin{align}
\frac 1 {h^2} ({q^i}^{(k+1)}-2{q^i}^{(k)}+{q^i}^{(k-1)})=V'({q^i}^{(k)})\,.
\end{align}
This is just what is wanted for the difference discrete counterpart of the 
equations in the continuous case.

Now we consider the difference discrete symplectic structure and its 
preserving  property. Taking the exterior derivative d on  $ L_{D(k)}$, 
we get
$$
\d L_{D(k)}= \frac{\partial L_{D(k)}}{\partial q^i_{(k)}} \d q^i_{(k)}
+\frac{\partial L_{D(k)}}{\partial q^i_{t(k)}}\d q^i_{t(k)}\,.
$$
By means of the modified Leibniz law with respect to $\Delta_t$ and 
introducing the DEL one-form
$$
\displaylines{\hspace*{8mm}
E_{D(k)}(q^i_{(k)}, q^j_{t (k)})
\hfill\cr\noa2
\hspace*{8mm}\qquad:\;=\Bigl\{\frac {\partial L_{D(k)}} 
{\partial q^i_{(k)}}-\Delta_t\Bigl(\frac {\partial L_{D(k-1)}} 
{\partial q^i_{t(k-1)}}\Bigr)\Bigr\}\d q^i_{(k)}\,,
\hfill{(33)}\cr}
$$
we have
\setcounter{equation}{33}
\begin{align}
\d L_{D(k)}=E_{D(k)}+\Delta_t\theta_{{\ssc D}(k)}\,,
\end{align}
where $\theta_{{\ssc D}(k)}$ is the discrete canonical one-form
\begin{align}
\theta_{{\ssc D}(k)}=\frac{\partial L_{D(k-1)}} {\partial q^i_{t(k-1)}}\d 
q^i_{(k)}\,,
\end{align}
and there exists the following discrete symplectic two-form on $T^*(M_{T_D}^n)$,
\begin{align}\alw
&\om_{{\ssc D}(k)}=\d\theta_{{\ssc D}(k)}
=\frac {\partial^2 L_{D(k-1)}}{\partial q^i_{t(k-1)} 
\partial q^j_{(k-1)}} \d q^j_{(k-1)} \wedge \d q^i_{(k)}
\nonumber\\\noa2
&\phantom{\om_{D(k)}=}+\frac {\partial^2 L_{D(k-1)}} {\partial q^i_{t(k-1)} 
\partial q^j_{t(k-1)}} \d q^j_{t(k-1)} \wedge \d q^i_{(k)}\,.
\end{align}
Now by virtue of the nilpotency of d on  $T^*(M_{T_D}^n)$, we get
\begin{align}
0=\d^2L_{D(k)}=\d E_{D(k)}+\Delta_t \om_{{\ssc D}(k)}\,.
\end{align}
Therefore, it is easy to see that if and only if the DEL one-form satisfies 
what is called the  DEL condition, i.e., it is closed
\begin{align}
\d E_{D(k)}=0\,,
\end{align}
then it gives rise to the discrete (difference) symplectic 
structure-preserving law,
\begin{align}
0=\Delta_t\omega_{D(k)}\,.
\end{align}

Similar to the continuous case, the null DEL 
one-form, which is corresponding to the DEL equation, is a special case
of the coboundary DEL one-forms,
\begin{align}
 E_{D(k)}=\d\alpha_{{\ssc D}(k)}(q^i_{(k)}, q^j_{t (k)})\,,
\end{align}
where $\alpha_{{\ssc D}(k)}(q^i_{(k)}, q^j_{t (k)})$ 
is an arbitrary function of 
$(q^i_{(k)}, q^j_{t (k)})$. Although they satisfy the DEL condition, 
this does not mean that all closed DEL one-forms are exact. In fact 
equation (34) shows that the EL one-forms are not exact in general 
since the canonical one-form $\theta_{{\ssc D}(k)}$ is not trivial. 
In addition,
$(q^i_{(k)}, q^j_{t (k)})$ are {\bfit not} in the solution space of the DEL
equations only rather they are still in the function space with
the DEL condition in general.  Therefore, this also means that the DDSSP 
law  holds in the function space with the DEL cohomology in general
rather than on the solution space only.

\subsect{The Discrete Multisymplectic Structure in
\newline\phantom{3.2\ \ }Discrete Lagrangian Field Theory }

We now study the discrete multisymplectic structure 
in discrete Lagrangian field theory. For the sake of simplicity, 
let us consider  the $1+1-d$ and $2-d$ cases in discrete 
Lagrangian field theory  (DLFT) for  a scalar field. Let $X^2$ with 
suitable signature of the 
metrics be the base manifold, $L^2$ a regular lattice with two-directions 
$x_{\mu}$, $\mu=1, 2$  on $X^2$, $M_D$ the discrete configuration space with 
$u^{(i,j)} \in M_D$. 

The difference discrete Lagrangian is denoted as
\begin{align}
L_D^{(i,j)}=L_D( u^{(i,j)} ,u_{\mu}^{(i,j)} )\,,
\end{align}
where
\begin{align*}\alw
&\Delta_1 u^{(i,j)} =\frac 1 h (u^{(i+1,j)} -u^{(i,j)} )\,,
\\\noa2 
&\Delta_2 u^{(i,j)} =\frac 1 h (u^{(i,j+1)} -u^{(i,j)} )\,.
\end{align*}
They are the bases of $T(M_D)$ and their duals $\d x^{\mu}=\d_L x^{\mu}$ are 
the bases of $T^*(M_D)$,  
$$
\d_L x^{\mu}(\partial_{\nu})=\delta_{\nu}^{\mu}\,.
$$

The action functional is given by
\begin{align}
S_D=\sum_{\{i,j\} \in {Z \times Z}}L_D(u^{(i,j)}, u_{\mu}^{(i,j)})\,.
\end{align}
Taking the variation of $S_D$ and regarding the differences as the entire
geometric objects, we get
$$
\delta S_D =\sum_{\{i,j\} \in {Z \times Z}}
\Bigl\{\frac {\partial L_D^{(i,j)}} {\partial u^{(i,j)}} \delta u^{(i,j)}
+\frac {\partial L_D^{(i,j)}} {\partial u_{\mu}^{(i,j)}} \delta u_{\mu}^{(i,j)}
\Bigr\}\,.
$$
Employing the modified Leibniz law, we have
\end{multicols}
\ruleup

\begin{align*}
&\Delta_1\Bigl(\frac{\partial L_D^{(i-1,j)}} {\partial u_{1}^{(k-1,l)}} 
\delta u^{(k,l)}\Bigr)
=\frac {\partial L_D^{(i,j)}} {\partial u_{1}^{(k,l)}} \delta u_{1}^{(k,l)}
+\Delta_1 \Bigl(\frac {\partial L_D^{(i-1,j)}} {\partial 
u_{1}^{(k-1,l)}}\Bigr) \delta u^{(k,l)}\,,
\\\noa2
&\Delta_2\Bigl(\frac {\partial L_D^{(i,j-1)}} {\partial u_{2}^{(k,l-1)}} 
\delta u^{(k,l)}\Bigr)
=\frac{\partial L_D^{(i,j)}}{\partial u_{2}^{(k,l)}} \delta u_{2}^{(k,l)}
+\Delta_2\Bigl(\frac {\partial L_D^{(i,j-1)}} 
{\partial u_{2}^{(k,l-1)}}\Bigr) \delta u^{(k,l)}\,.
\end{align*}

\ruledown
\begin{multicols}{2}
\noindent
Assuming that $\delta u^{(k,l)}$s vanish at infinity, it follows the DEL
equations
\begin{align}
\frac {\partial L_D^{(i,j)}} {\partial u^{(k,l)}} 
-\Delta_1 \Bigl(\frac {\partial L_D^{(i-1,j)}} 
{\partial u_{1}^{(k-1,l)}}\Bigr) 
-\Delta_2\Bigl(\frac {\partial L_D^{(i,j-1)}} 
{\partial u_{2}^{(k,l-1)}}\Bigr) =0\,.
\end{align}

Let us  consider an example with the discrete Lagrangian
\begin{align}
L_D^{(i,j)}( u^{(i,j)} ,u_{\mu}^{(i,j)} )
=\frac 1 2 (\Delta_{\mu} u^{(i,j)})^2-V( u^{(i,j)})\,.
\end{align}
The DDVP gives the DEL equations
$$
\Delta_1(\Delta_1  u^{(i-1,j)})+\Delta_2(\Delta_2  u^{(i,j-1)})
-V'( u^{(i,j)})=0\,,
$$
i.e.,
$$
\displaylines{\hspace*{6mm}
\frac1 {h_1^2} (u^{(i+1,j)}-2u^{(i,j)}+u^{(i-1,j)})
+\frac 1 {h_2^2} (u^{(i,j+1)}
\hfill\cr\noa2
\hspace*{6mm}\phantom{\frac1 {h_1^2} (u^{(i+1,j)}}-2u^{(i,j)}+u^{(i,j-1)})
=V'(u^{(i,j)})\,.
\hfill{(45)}\cr}
$$
This is also what is wanted for the difference discrete counterpart of the
relevant PDEs in the continuous limit.

We now consider the multisymplectic properties of the DLFT.
Taking exterior derivative $\d \in T^*(M_D)$ of $L_D^{(i,j)}$ and 
making use of the  modified Leibniz law, we get
\setcounter{equation}{45}
\begin{align}
\d L_D^{(i,j)}=E_D^{(i,j)}(u^{(i,j)}, u_{\mu}^{(i,j)})
+\Delta_{\mu} \theta^{\mu (i,j)}\,,
\end{align}
where $E_D^{(i,j)}$ is the DEL one-form defined by
$$
\displaylines{
E_D^{(i,j)}(u^{(i,j)}, u_{\mu}^{(i,j)}):\;=
\Bigl\{\frac {\partial L_D^{(i,j)}} {\partial u^{(k,l)}} 
-\Delta_1 \Bigl(\frac {\partial L_D^{(i-1,j)}} 
{\partial u_{1}^{(k-1,l)}}\Bigr) 
\hfill\cr\noa2
\phantom{E_D^{(i,j)}(u^{(i,j)}, u_{\mu}^{(i,j)}):=}
-\Delta_2\Bigl(\frac {\partial L_D^{(i,j-1)}} {\partial u_{2}^{(k,l-1)}}
\Bigr) \Bigr\}\d u^{(k,l)}\,,
\hfill{(47)}\cr}
$$
and $\theta^{\mu (i,j)}$ are two Cartan one-forms,
\setcounter{equation}{47}
\begin{align}\alw
&\theta^{1 (i,j)}=\frac {\partial L_D^{(i-1,j)}} {\partial u_{1}^{(k-1,l)}} 
\d u^{(k,l)}\,,
\nonumber\\\noa2
&\theta^{2 (i,j)}=\frac{\partial L_D^{(i,j-1)}} {\partial u_{2}^{(k,l-1)}} 
\d u^{(k,l)}\,.
\end{align}
It is easy to see that there exist two symplectic two-forms on  $T^*(M_D)$,
\begin{align}
\om^{\mu (i,j)}=\d\theta^{\mu (i,j)},\qquad  \mu=1,2\,.
\end{align}
The equation $\d^2L_D^{(i,j)}=0$  on  $T^*(M_D)$ leads to the conservation 
law or the divergence free equation  of $\om^{\mu (i,j)}$,
\begin{align}
\Delta_{\mu}\om^{\mu(i,j)}=0\,,
\end{align}
if and only if the DEL one-form satisfies the DEL condition, 
i.e., it is closed 
\begin{align}
\d E_D^{(i,j)}=0\,.
\end{align}

Similar to the continuous case,  the null DEL one-form is  
corresponding to the 
DEL equations and it is a special case of coboundary DEL one-forms
\begin{align}
 E_D^{(i,j)}=\d\alpha_{\ssc D}^{(i,j)}\,,
\end{align}
where $\alpha_{\ssc D}^{(i,j)}$  is an arbitrary function on $T^*M_D$. 
Although they
satisfy the DEL condition, it does not mean that all closed DEL one-forms
are exact. As  a matter of fact, from Eq.~(46) it is easy
to see that the EL one-forms are not exact in general since the two canonical
one-forms $\theta^{\mu (i,j)}$  $(\mu=1,2) $ are not trivial. 
In addition, this indicates that the variables $u^{(k,l)}$s
are still in the function space in general rather than the ones in the 
solution space only. Consequently, this also means that the difference 
discrete multisymplectic structure-preserving law  holds in the 
function space with the closed DEL condition in general rather than 
in the solution space only.

It should be pointed out that the scenario of the approach 
can be straightforwardly  generalized to higher-dimensional cases of 
$X^{1,n-1}$ and $X^{n}$.

\sect{Remarks}
\vspace*{-0.15cm}

A few remarks are in order.

1) The approach presented in this paper, which may call the EL cohomology
approach, to the symplectic and multisymplectic geometry and their 
difference discrete versions in 
the Lagrangian formalisms is more or less different
from other approaches.\wen{3,4} The EL and the DEL cohomological 
concepts and relevant content such as the EL and DEL one-forms,
the null EL and DEL one-forms, the coboundary
EL and DEL one-forms as well as the  EL and the DEL conditions have been 
introduced and they have played very crucial roles in each case to show 
that the symplectic and  multisymplectic
preserving properties and their difference discrete versions are 
in the function space with the closed EL/DEL condition in 
general rather than in the solution space only.

It has been mentioned  that the EL and DEL cohomology in relevant case is not
trivial and it is very closely related to the symplectic and multisymplectic
structures as well as their discrete versions. 
Needless to say, the content and the role of the EL cohomology 
in each case should be further studied and some issues are under 
investigation.\wen{8} 

2) The difference discrete variational formalism presented here is also 
different from the one by Veselov.\wen{9,10} We have emphasized that the 
difference as discrete derivative is an entire geometric object. 
This is obvious and natural  from the view point
of NCDC. The continuous limits of the results given here are correct as well.

3) The NCDC on the regular lattices are employed in our approach. Since the
differences do not satisfy the ordinary Leibniz law, in order to study the
symplectic and multisymplectic geometry in these difference discrete
systems it is natural and meaningful to make use of the NCDC.

4) The approach presented here can be generalized to the case that the 
configuration space is also discretized. This  is closely related to 
the case of difference discrete phase space approach to the finite 
dimensional  systems with separable Hamiltonian.\wen{5,6}

5) It should be mentioned that the approach with the EL 
cohomological concepts can also directly be applied to the PDEs.

Let us for example consider the following type of equations\wen{3,11}  
and make use of the same notations in Refs~[3] and [11], 
\begin{align}
Kz_{x_1}+Lz_{x_2}=\nabla_zS(z)\,.
\end{align}
Introducing the EL one-form
\begin{align}
E(z,z_{x_1},z_{x_2}):\;=\d z^T\{Kz_{x_1}+Lz_{x_2}-\nabla_zS(z) \}\,,
\end{align}
it is easy to see that the null EL one-form gives rise to
Eq.~(54) and it is a special case of the coboundary EL one-forms,
\begin{align}
E(z, z_{x_1},z_{x_2})=\d\alpha(z, z_{x_1},z_{x_2})\,,
\end{align}
where $\alpha(z, z_{x_1},z_{x_2})$ is an arbitrary function of 
$(z, z_{x_1},z_{x_2})$.

Now by taking the exterior derivative d of the EL one-form, it is 
straightforward to prove that
$$
\d E(z, z_{x_1},z_{x_2})=\frac12\partial_{x1}(\d z^T \wedge K\d z)
+\frac12\partial_{x_2}(\d z^T \wedge L\d z).
$$
This means that the following MSSP equation
\begin{align}
\partial_{x_1} \omega +\partial_{x_2} \tau=0
\end{align}
holds, where 
$$
\omega=\d z^T \wedge K\d z, \qquad \tau= \d z^T \wedge L\d z\,,
$$
if and only if the EL one-form is closed, i.e.,
\begin{align}
\d E(z, z_{x_1},z_{x_2})=0\,.
\end{align}

It should be mentioned that first from the definition of the EL one-form, 
it is not trivial in general since the first two terms in the 
definition are the 
canonical one-forms which are not trivial so that the EL cohomology is not
trivial. Secondly, the multisymplectic preserving equation derived here is
not dependent on the type of equations but can be applied to the equations 
so that it  holds not only in the solution space of the equations but also 
in the function space relevant to the cohomology in general as well.

6) In principle, the cohomological scenario presented here should be 
available to not only to PDEs but also to ODEs and numerical schemes as well.

7) Finally, it should be pointed out that there exist lots of other 
problems to be studied.

\renewcommand\theequation{A\arabic{equation}}
\setcounter{equation}{0}

\nonumsect{Appendix}
\vspace*{-0.15cm}

In this appendix we briefly recall some content of NCDC on lattice.\wen{5-7}

\vspace*{0.15cm}

\noindent
{\bfit A.1 \  An NCDC on an Abelian Discrete Group}

Let $G$ be an  Abelian discrete group with a generator $t$, $A$ the  algebra 
of  complex valued functions on $G$.

The left and  right multiplications of a generator of $G$ on its element are
commute  to each other  since $G$ is  Abelian.  Let us  introduce right
action on $A$ that is given by
\begin{align}
R_t f(a)=f(a \cdot t)\,,
\end{align}
where  $f \in  A$,  $a \in  G$, $t$ the generator and ``$\cdot$''  the group
multiplication.

Let   $V$  be   the   space  of   vector  fields,
$$
V=\text{span}   \{  \partial_t\}\,,
$$
where $\partial_t$  is the derivative with respect to the generator
$t$ given by
\begin{align}
(\partial_t f)(a)\equiv R_t f(a)-f(a) = f(a\cdot t)-f(a)\,.
\end{align}
The dual space of $V$, the space of one-form, is 
$\Omega ^1=\text{span} \{ \chi^t\}$  that is dual to $V$,
\begin{align}
\chi^t(\partial_t)=1\,.
\end{align}
The  whole differential  algebra  $\Omega^*$  can also  be defined  as
$\Omega^*= \bigoplus_{n=0,1} \Omega^n $ with $A=\Omega^0$.

Let us define the exterior differentiation in 
$$\Omega^*\d : ~ \Omega^0 \rightarrow \Omega^{1}\,.$$
It acts on a zero-form  $\omega^0 = A $ is as follows:
\begin{align}
\d f=\partial_tf \chi^t\in \Omega^{1}\,.
\end{align}
Now, the following theorem can straightforwardly be proved.

\vspace*{0.15cm}

\noindent
{\bfit Theorem\ \ }The exterior differentiation d satisfies
$$
\displaylines{\hspace*{10mm}
(\text{a}) ~ (\d f)(v)=v(f),\quad v\in V,\quad f\in \Omega^0\,,
\hfill\cr\noa2
\hspace*{10mm}(\text{b}) ~ \d^{2}=0\,,
\hfill\cr\noa2
\hspace*{10mm}(\text{c}) ~ \d(\omega\otimes\omega')=\d\omega\otimes\omega'
+(-1)^{\text{deg}\,\omega}\omega\otimes \d\omega'\,,
\hfill{\text{(A5)}}\cr}
$$
if and only if
\setcounter{equation}{5}
\begin{align}\alw
&(\text{i})\quad \d\chi^{t}=0 \,, 
\nonumber\\\noa2
&(\text{ii}) \quad \chi^{t}f=(R_{t}f) \chi^{t}\,.
\end{align}

As  was shown  here, in  order to  establish a  well-defined differential
algebra, it  is necessary  and sufficient to  introduce the noncommutative
property of the multiplication between function and one-form.

The conjugation * on the whole differential algebra $ \Omega ^{*} $
and metric on discrete Abelian group can also be defined.

\vspace*{0.15cm}

\noindent
{\bfit A.2 \  An NCDC on Regular Lattice} 

Let us consider the discrete translation group
$ G^{m}= \otimes_{i=1}^m G^i  $, $A$ the function space on $G^m$ and a 
regular lattice with equal  spacing $L^m$ on an $m$-dimensional  space $R^m$. 
Here $G^i$ the $i$-th discrete translation group with one generator acting on 
one-dimensional space with coordinate $q$ in such a way
\begin{align}
R_{q^i}: q^i_n\rightarrow q^i_{n+1}=q^i_n+h,\qquad h\in R_+\,,
\end{align}
$R_{q^i}$ the  discrete translation operation of the  group $G^i$ and it maps
$q^i_n$ of  $n$-th size of $q^i$ to the one $q^i_{n+1}$ at  {$(n+1)$-th} size, 
$h$ the discrete translation step-length  and $R_+$ the positive real  
number. It is easy to see that  the action of $G^i$ on $i$-th one-dimensional  
space $R^1$ generates the $i$-th  chain $L^i$, $i=1, \ldots , m,$
with  equal spacing $h$.  Similarly, the  regular lattice $L^m$  with equal
spacing  $h$  is generated  by  $G^m$ acting  on  $R^m$. Since  there is  a
one-to-one correspondence between sizes on $L^i$ and elements of $G^i$, one
may  simply regard  $L^i$ as  $G^i$. For  the same  reason, one  may simply
regard $L^m$ as $G^m$. 

On the sizes of the regular lattice $L^{m}$, there are discrete coordinates
$q^i_n$, $i=1, \ldots , m$.  There is a set of   generators in
the  discrete  translation group  $G^{m}$  acting on $L^{m}$ in such a way
\begin{align}
R_{q^i}:  ~q^{i}_{n} \to q^{i}_{n+1}, \qquad i=1, \ldots, m\,.
\end{align}
With  respect to  the generators  there is  a set of independent 
 derivatives $\partial_{q^i}$ on $f_{n}(q^i)=f(q^i_n) \in  A$. They should
be defined  as the correspondent differences of  the functions valued 
at two nearest sizes, i.e.,
\begin{align}
&\partial_{q^i}f(q^i_n)=\Delta_{q^i}f(q^i_n)=\frac{1}{h}[
(R_{q^i}-i\d)f(q^i_n)]
\nonumber\\\noa2
&\phantom{\partial_{q^i}f(q^i_n)}=\frac{1}{h}[f(q^i_{n+1})-f(q^i_n)]\,.
\end{align}

The differential one-form is defined by
\begin{align}
\d f=\partial_{q^i} f\d q^i=\Delta_{q^i}f\d q^i, \qquad f \in A\,.
\end{align}
The  two-forms  and the  whole  differential  algebra
$\Omega^*$ can  also be defined. Here  d is the exterior differentiation.
Similarly, the following theorem can be proved for d.

\vspace*{0.15cm}

\noindent
{\bfit  Theorem\ \ } d is  nilpotent  and  satisfies the  Leibniz rule, i.e.,
$$
\displaylines{\hspace*{10mm}
\d^2=0\,,
\hfill\cr\noa2
\hspace*{10mm}\d(\omega\wedge\omega')=\d\omega\wedge\omega'
\hfill\cr\noa2
\hspace*{10mm}\phantom{\d(\omega\wedge\omega')=}
+(-1)^{\text{deg}\,\omega}\omega\wedge \d\omega',
~~\omega,\omega' \in \Omega^*\,,
\hfill{\text{(A11)}}\cr}
$$
if and only if
\setcounter{equation}{11}
\begin{align}
f(q^{i}+h)\d q^{i}=\d q^{i}f(q^{i})\,.
\end{align}
This gives
$$
q^{i}\d q^{i}-\d q^{i}q^{i}=-h\d q^{i}\,.
$$
The  above two  equations show  the noncommutative properties  between the
functions   (including    the   coordinates)    and   differential   forms.

From  these   properties,  it   follows  the  modified   Leibniz  rule  for
derivatives,
\begin{align}
\Delta_{q^{i}}(fg)=\Delta_{q^{i}}f \cdot g + \{R_{q^{i}}f\} 
\cdot \Delta_{q^{i}}g\,.   
\end{align}

It should be noted that the definitions and relations given above for the 
NCDC on the regular lattice  $L^{m}$ are at least formally very similar 
to the ones in the ordinary commutative differential calculus (CDC) on  
$R^{m}$. The differences between the two cases are commutative or not. 

The Hodge $*$ operator and the co-differentiation operator 
$$
\delta_L: ~ \Omega^k \rightarrow \Omega^{k-1}
$$ 
on the regular
lattice  $L^{m}$ can also be defined similarly to the ones on   $R^{m}$. 
Consequently, the Laplacian on the lattice  $L^{m}$ may also be given by
\begin{align}
\Delta_L=\d \delta_L+\delta_L \d\,.
\end{align}
It is in fact the discrete counterpart of the Laplacian $\Delta$ on  $R^{m}$. 
For other objects and/or properties on  $R^{m}$, there may have the discrete 
counterparts on  $L^{m}$ as well. For example, the null-divergence 
equation of a form $\om $ on $R^m$ reads 
\begin{align}
\delta \alpha =0\,.
\end{align}
Its counterpart on the lattice $L^m$ is simply
\begin{align}
\delta_L \alpha_L =0\,.
\end{align}
This is the forward difference form of null-divergence equation.

In the case of $L^{1,m} \in R^{1,m}$ with Lorentz signature, these equations 
become the conservation law of $\alpha$ and its difference form of 
$\alpha_{\ssc L}$. This is available not only for the symplectic geometry and 
symplectic algorithms but also the multisymplectic geometry and 
multisymplectic algorithms as well. It should be emphasized that  for the 
discrete counterparts on the lattice, they obey the NCDC on the lattice 
$L^{m}$ rather than the CDC on $R^{m}$. This is the most important point.
\end{multicols}

\refrule

\begin{multicols}{2}

\end{multicols}

\vfill
\end{document}